# Indices to detect Hopf bifurcation in Induction motor drives

Krishnendu Chakrabarty
Kalyani Government Engineering College
Kalyani-741235, India
chakrabarty40@rediffmail.com

Urmila Kar
National Institute of Technical Teachers' Training and Research
Kolkata-700106, India
urmilakar@rediffmail.com

*Abstract*— **The loss of stability of induction motor controlled by Indirect Field Oriented Control (IFOC) is a matter of great concern of operators and design engineers. This paper reports indices to detect and predict stability problem such as system oscillations. Oscillations as a result of loss of stability, due to Hopf bifurcation, for different parameter values of IFOC motor are studied using the proposed indices.**

## I. INTRODUCTION

Nowadays, induction motors are extensively used for household and industrial applications for its relatively rugged and inexpensive structure. Therefore, much attention is given to design controllers to control the speed of the motor for various applications. The vector control of AC drives has been used widely in high performance control. Indirect field oriented control (IFOC) is one of the most effective vector control method to control induction motor due to the simplicity of design, better stability and high performance.

The effect of error in estimation of rotor resistance of IFOC motor in the light of stability has been investigated [1]. It has been previously shown that the IFOC based speed control of induction motors with constant speed are globally stable [2]. An analysis of saddle-node and Hopf bifurcation in IFOC drives due to errors in the estimate of the rotor time constant provides a guideline for setting the gains of PI speed controller in order to avoid Hopf bifurcation [3].Stable drives are required for conventional applications. Hopf bifurcation in IFOC drives with estimation error of rotor time constant as bifurcation parameter and condition of Hopf bifurcation have been analyzed [4]. Bifurcation causes instability in IFOC motors. A criteria for occurrence of Hopf bifurcation in IFOC motor has been reported. This criteria is used to find bifurcation surface in IFOC motors [5]. Voltage collapse or oscillation in power system is a point of major concern to the researchers. There are evidences of Hopf bifurcation in power system. An index to identify Hopf bifurcation in power system has been proposed [6].

For stable operation of IFOC motor, it is very essential to know the value of the parameter or region in the parameter space where motor becomes unstable due to Hopf bifurcation. In this paper, two indices to identify Hopf bifurcation has been proposed. Both of them show a smooth and predictable shape and can be readily computed so that the prediction of instability due to Hopf bifurcation can be done online.

The paper has been organized as follows: The mathematical model of IFOC induction motor is obtained firstly in section II. The Hopf bifurcation and bifurcation indices are explained in detail in section III. The dynamics of the system are given in section IV. The bifurcation behavior of the system along with application of indices are presented in section V. The conclusion is given in section VI.

## II. MATHEMATICAL MODEL OF THE SYSTEM

The nonlinear dynamical model of a current fed induction machine is selected for bifurcation analysis. Stator currents are the input to the system. Rotor equations of the induction motor expressed in a reference frame rotating at synchronous speed are given as follows:

$$\begin{aligned} &R_r i_{dr} - \omega_{sl}\psi_{qr} + \dot{\psi}_{dr} = 0 \\ &R_r i_{qr} + \omega_{sl}\psi_{dr} + \dot{\psi}_{qr} = 0 \\ &\psi_{qr} = L_r i_{qr} + L_m i_{qs} \\ &\psi_{dr} = L_r + L_m i_{ds} \end{aligned} \tag{1}$$

If the machine has distributed windings and sinusoidal MMF, then only d-q variables are taken in torque equation. The electrical torque are expressed as

$$T_e = \frac{3}{2} P \frac{L_m}{L_r} (\psi_{dr} i_{qs} - \psi_{qr} i_{ds}) \tag{2}$$

The mechanical equation is given by

$$\dot{\omega} = \frac{1}{J}(T_e - T_l - \beta\omega) \tag{3}$$

From equation (1), (2) and (3), the simplified model of the induction motor can be expressed as:

$$\dot{\psi}_{qr} = -\frac{R_r}{L_r}\psi_{qr} - \omega_{sl}\psi_{dr} + \frac{L_m}{L_r}R_r i_{qs}$$

$$\dot{\psi}_{dr} = -\frac{R_r}{L_r}\psi_{dr} + \omega_{sl}\psi_{qr} + \frac{L_m}{L_r}R_r i_{ds} \qquad (4)$$

$$\dot{\omega} = -\frac{\beta}{J}\omega + \frac{1}{J}\left[\frac{3}{2}\frac{L_m}{L_r}P(\psi_{dr}i_{qs} - \psi_{qr}i_{ds}) - T_l\right]$$

In order to form the state-space equation, the following states, inputs and constants are defined:

$$x_1 = \psi_{qr}, x_2 = \psi_{dr}, u_1 = \omega_{sl}, u_2 = i_{ds}, u_3 = i_{qs}$$

$$c_1 = \frac{R_r}{L_r}, c_2 = \frac{L_m}{L_r}R_r, c_3 = \frac{\beta}{J}, c_4 = \frac{1}{J},$$

$$c_5 = \frac{3}{2}\frac{L_m}{L_r}P \qquad (5)$$

Including (5) into (4), the state–space model can be written as :

$$\dot{x}_1 = -c_1 x_1 + c_2 u_3 - u_1 x_2$$

$$\dot{x}_2 = -c_1 x_2 + c_2 u_2 + u_1 x_1 \qquad (6)$$

$$\dot{\omega} = -c_3\omega + c_4\left[c_5(x_2 u_3 - x_1 u_2) - T_l\right]$$

Amongst the constants defined in (5), $c_1$ represents the inverse of the rotor time constants for the fundamental frequency and it has influence on the performance of the motor. To evaluate the impact of detuning of this parameter, a new constant is included and defined as the ratio between estimated and real rotor time constant

$$k = \frac{\hat{c}_1}{c_1}. \qquad (7)$$

As per control scheme of indirect field oriented control of induction motor, the controller equations are presented as:

$$u_1 = \hat{c}_1\frac{u_3}{u_2}$$

$$u_2 = u_2^0 \qquad (8)$$

$$u_3 = k_p(\omega^* - \omega) + k_i\int_0^t (\omega^*(\zeta) - \omega(\zeta))d\zeta$$

To form closed loop equations, controller equations in (8) are included in machine system of (6) and two new state variables are defined as:

$$x_3 = \omega^* - \omega \qquad \text{and} \qquad x_4 = u_3 \qquad (9)$$

Including (7)-(9) in (6), the state –variable equations of the motor for the indirect field oriented induction motor control can be expressed as:

$$\dot{x}_1 = -c_1 x_1 + c_2 x_4 - \frac{kc_1}{u_2^0}x_2 x_4$$

$$\dot{x}_2 = -c_1 x_2 + c_2 u_2^0 + \frac{kc_1}{u_2^0}x_1 x_4 \qquad (10)$$

$$\dot{x}_3 = -c_3 x_3 - c_4\left[c_5(x_2 x_4 - x_1 u_2^0) - T_l - \frac{c_3}{c_4}\omega^*\right]$$

$$\dot{x}_4 = (k_i - k_p c_3)x_3 - k_p c_4\left[c_5(x_2 x_4 - x_1 u_2^0) - T_l \frac{c_3}{c_4}\omega^*\right]$$

where $R_r$ is rotor resistance, $L_r$ is rotor self inductance, $L_m$ is the mutual inductance, P is number of pole pairs, $\omega_{sl}$ is slip angular speed, J is inertia coefficient, $T_l$ is load torque, $\psi_{qr}$ and $\psi_{dr}$ are the quadrature and direct axis flux linkage of the rotor, $\omega$ is the rotor angular speed, $\beta$ is friction coefficient, $\omega^*$ is reference speed, $u_2^0$ is the constant reference for the rotor flux magnitude, $i_{ds}$ is the direct axis stator current and $i_{qs}$ is the quadrature axis stator current.

The solution of the following algebraic equations obtained from (10) will give the equilibria of the system .

$$-c_1 x_1 + c_2 x_4 - \frac{kc_1}{u_2^0}x_2 x_4 = 0$$

$$-c_1 x_2 + c_2 u_2^0 + \frac{kc_1}{u_2^0}x_1 x_4 = 0 \qquad [11]$$

$$-c_3 x_3 - c_4\left[c_5(x_2 x_4 - x_1 u_2^0) - T_l - \frac{c_3}{c_4}\omega^*\right] = 0$$

$$(k_i - k_p c_3)x_3 - k_p c_4\left[c_5(x_2 x_4 - x_1 u_2^0) - T_l \frac{c_3}{c_4}\omega^*\right] = 0$$

Linearizing the system at the fixed point, the Jacobian matrix is obtained as follows:

$$J = \begin{bmatrix} -c_1 & -\frac{kc_1}{u_2^0}x_4 & 0 & c_2 - \frac{kc_1}{u_2^0}x_2 \\ \frac{kc_1}{u_2^0}x_4 & -c_1 & 0 & \frac{kc_1}{u_2^0}x_1 \\ c_4 c_5 u_2^0 & -c_4 c_5 x_4 & -c_3 & -c_2 c_5 x_2 \\ k_p c_4 c_5 u_2^0 & -k_p c_4 c_5 x_4 & k_i - k_p c_3 & -k_p c_4 c_5 x_2 \end{bmatrix} \qquad (12)$$

The eigen values of the system can be obtained from

$$|\lambda I - J| = 0 \qquad (13)$$

## III. Hopf Bifurcation and Hopf Bifurcation Indices

The term Hopf bifurcation refers to the local birth of a periodic solution from an equilibrium point as parameter of the system crosses a critical value. A Hopf bifurcation occurs when a complex conjugate pair of eigen values of the Jacobian of the system at a fixed point becomes purely imaginary i.e. real part of the eigen values becomes zero.
There are three indices to identify Hopf bifurcation.
Index1:
For the Jacobian matrix J, a complex pair of eigen value can be written as,

$$J[v_r \pm jv_i] = (a \pm jb)[v_r \pm jv_i] \qquad (14)$$

where a and b are the real and imaginary parts of the eigen value respectively. $v_r \pm iv_i$ are the associated eigen vectors. If real and imaginary parts are separated, the following relations are obtained:

$$(J - aI)v_r + bv_i = 0$$

$$(J - aI)v_i - bv_r = 0$$

$$\Rightarrow \begin{bmatrix} J - aI & bI \\ -bI & J - aI \end{bmatrix} \begin{bmatrix} v_r \\ v_i \end{bmatrix} \qquad (15)$$

$$\Rightarrow \left[ \begin{bmatrix} J & bI \\ -bI & J \end{bmatrix} - aI_{8x8} \right] \begin{bmatrix} v_r \\ v_i \end{bmatrix}$$

$$J_m = \begin{bmatrix} J & bI \\ -bI & J \end{bmatrix}$$

Where, I is an identity matrix having 4x4 dimensions.
As proposed by [6], the minimum singular value of the matrix $J_m$ can be used to detect Hopf bifurcation.

Hence, Index1= $\sigma_{\min}(J_m)$ (16)

Index2:
The Schur's decomposition of a matrix $J_m$ has a diagonal formed by real parts of eigen values of $J_m$.

Another index can be defined by

$$\text{Index2} = \sqrt[n]{\prod_{K=1}^{n} |S_{KK}|} \qquad (17)$$

Where $S_{kk}$ is diagonal elements of Schur($J_m$).

## IV. Dynamics of the System

The dynamics of the IFOC induction motor are studied with the help of computer simulation. The simulation are carried out with the following parameter values: $c_1$=13.67, $c_2$=1.56, $c_3$=0.59, $c_4$=1176, $c_5$=2.86, $k_p$=0.001, $k_i$=0.5, $w_{ref}$=181.1 rad/s.

Fig. 1 (a) and 2(b) show the period-1 limit cycle of IFOC induction motor. Limit cycles with period-2, period-3 and period-4 are shown in Fig. 2, 3& 4 respectively. The periodicity of the limit cycle is determined by the number of loops in the phase plane or the repetition of peaks in the time plot of a state variable. The presence of infinite loops in the phase plane or non repetition of peaks in the time plot indicates chaos. The chaotic attractor is shown in Fig. 5.

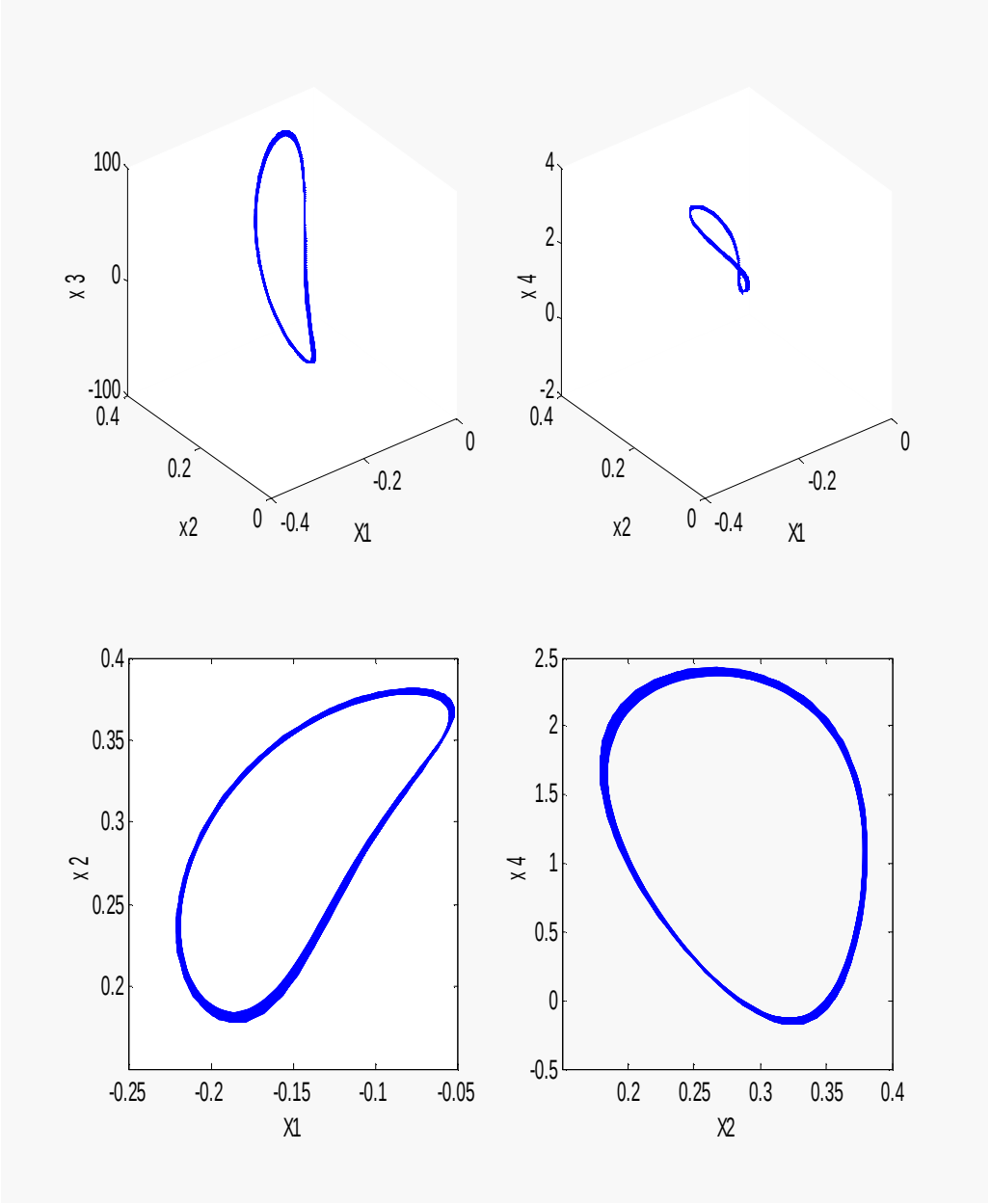

(a)

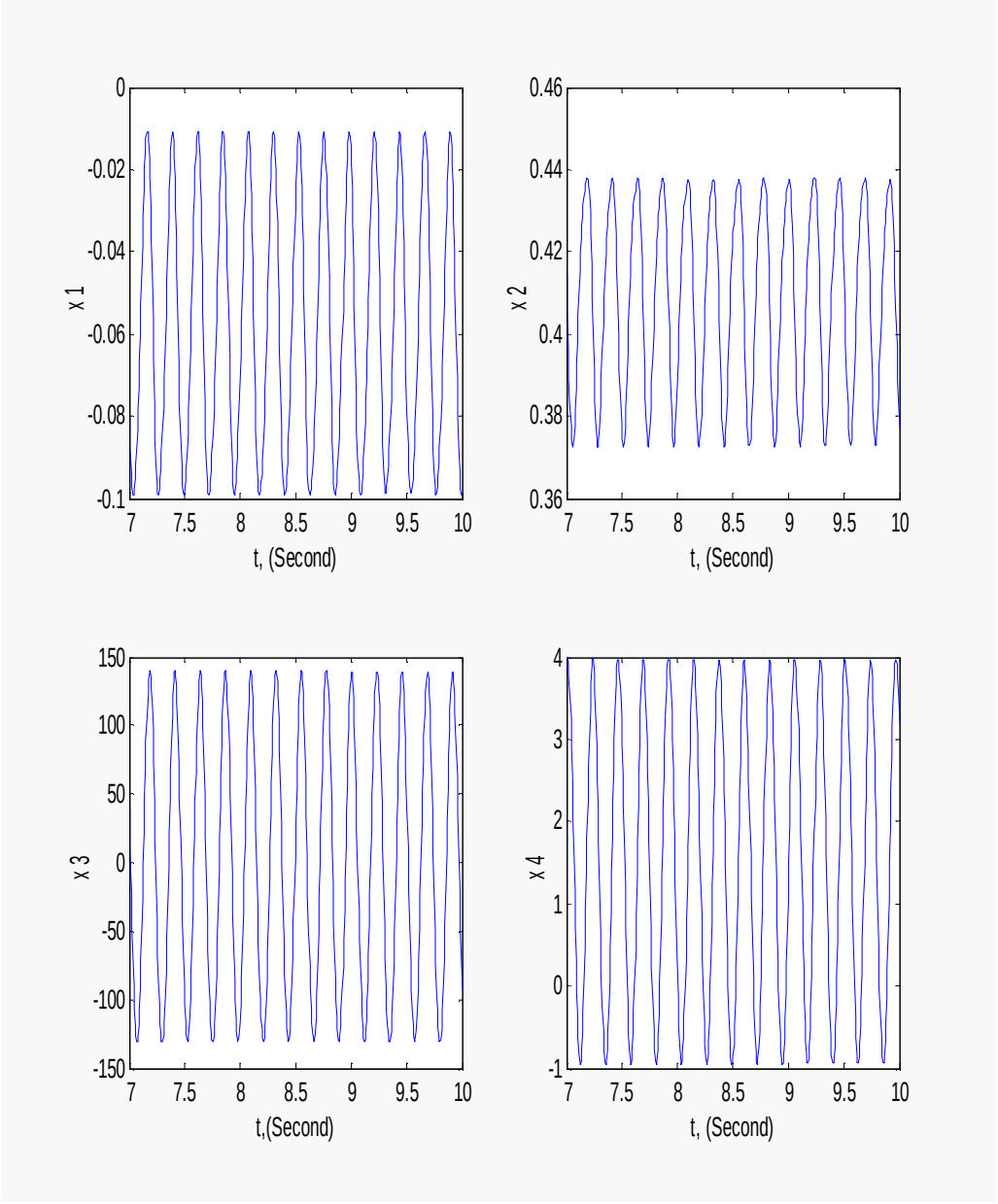

(b)

Fig. 1 (a) Phase plot at K=1.5 and Tl=2.3

(b) Time plot at K=1.5 and Tl=2.3

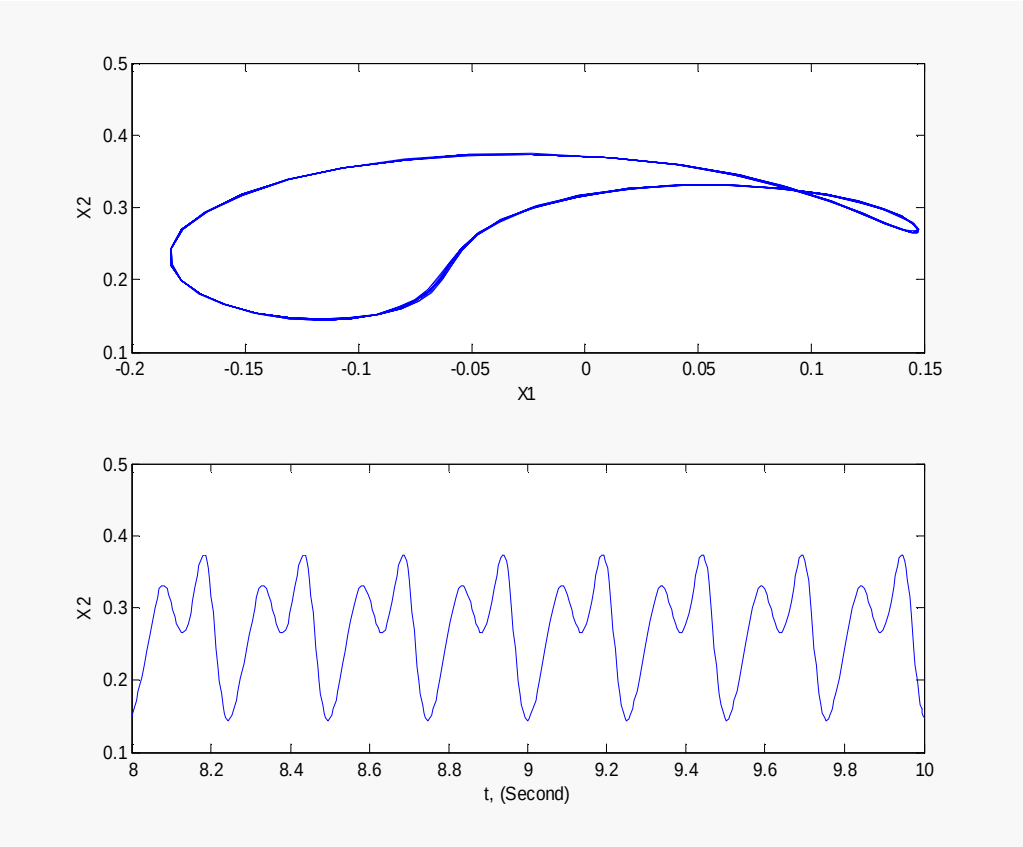

Fig. 2. Phase & Time plot at K=2.5 and Tl=0.5

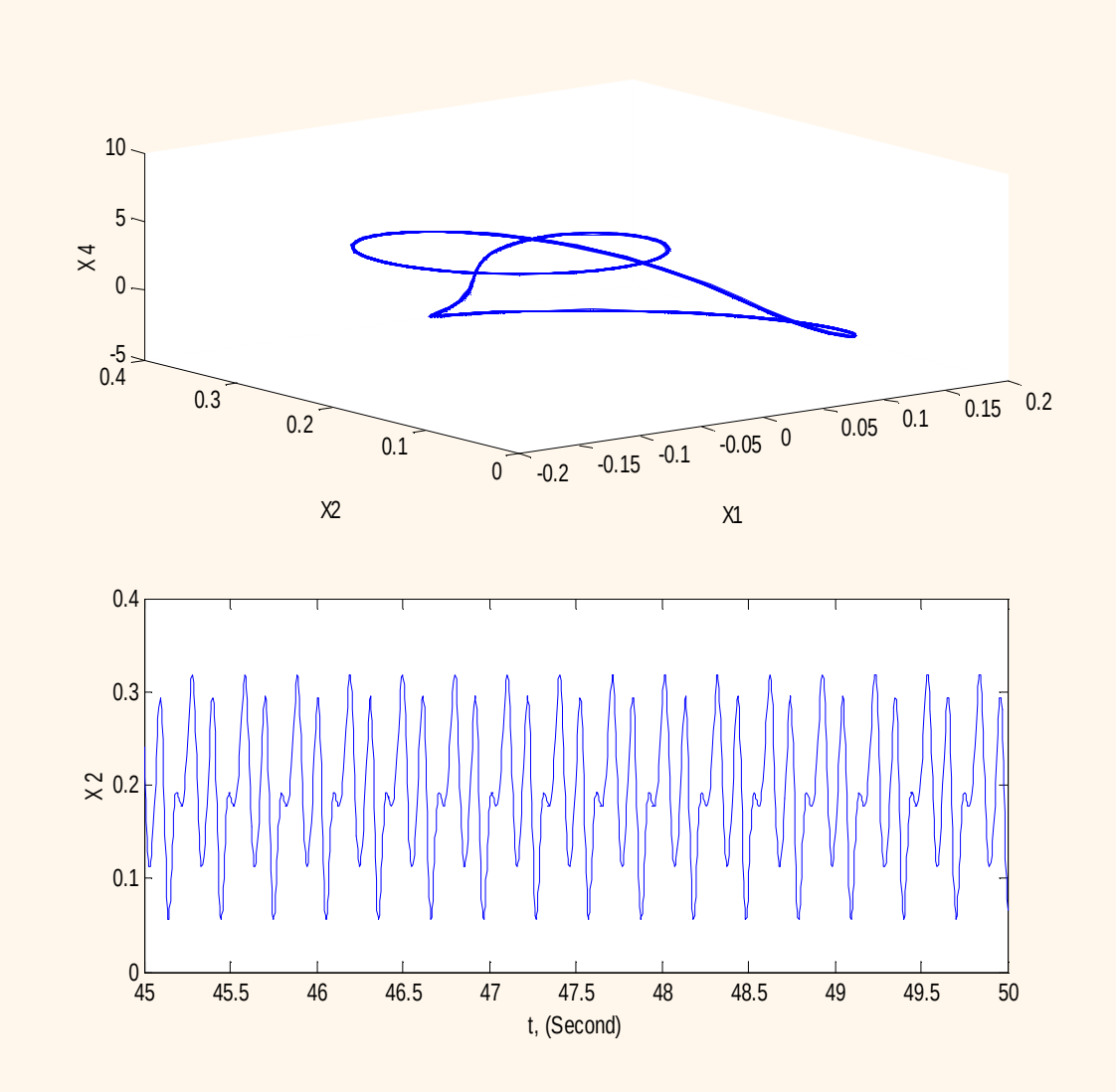

Fig. 3. Phase and time plot showing period-3 at K=3.25 and $T_l$=0.5

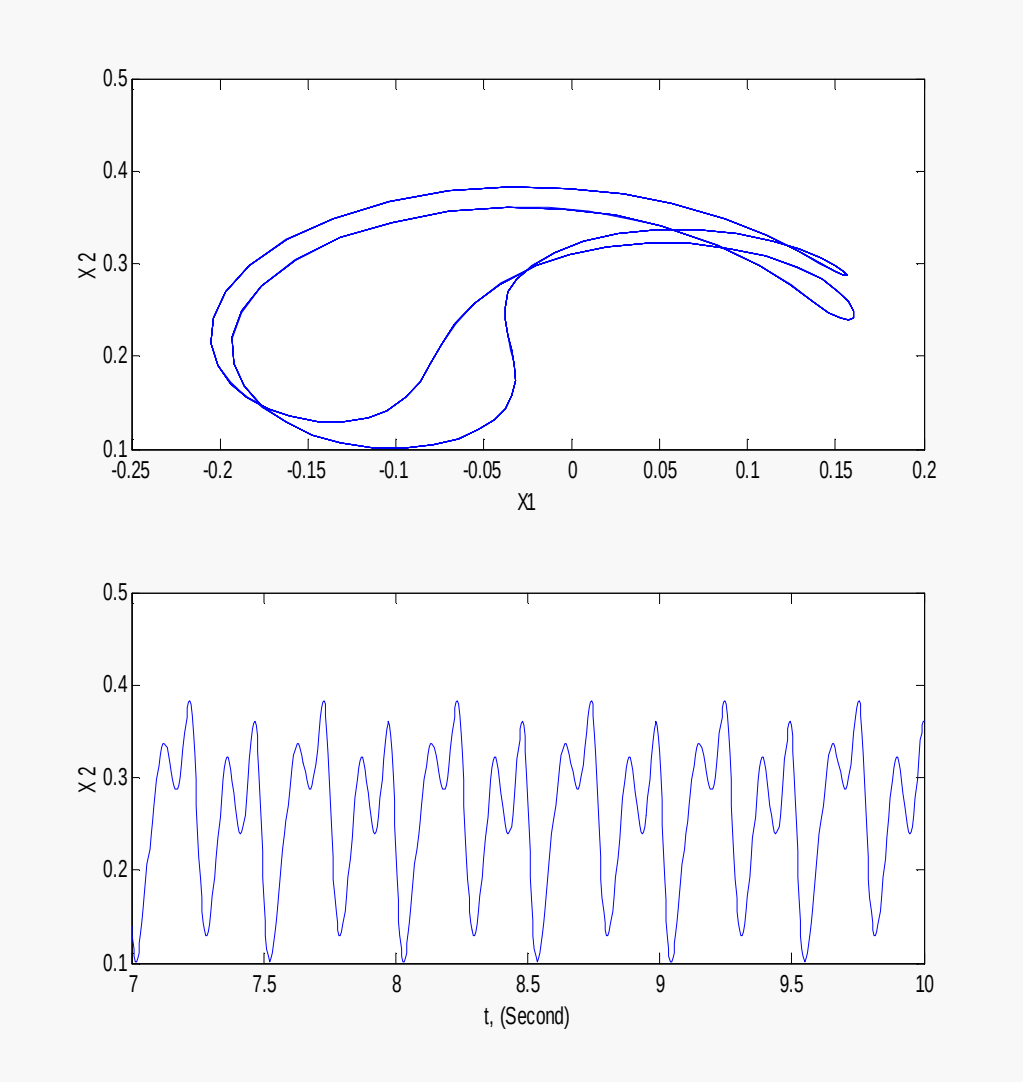

Fig. . 4. Phase & Time plot at K=2.8 and $T_l$=0.5

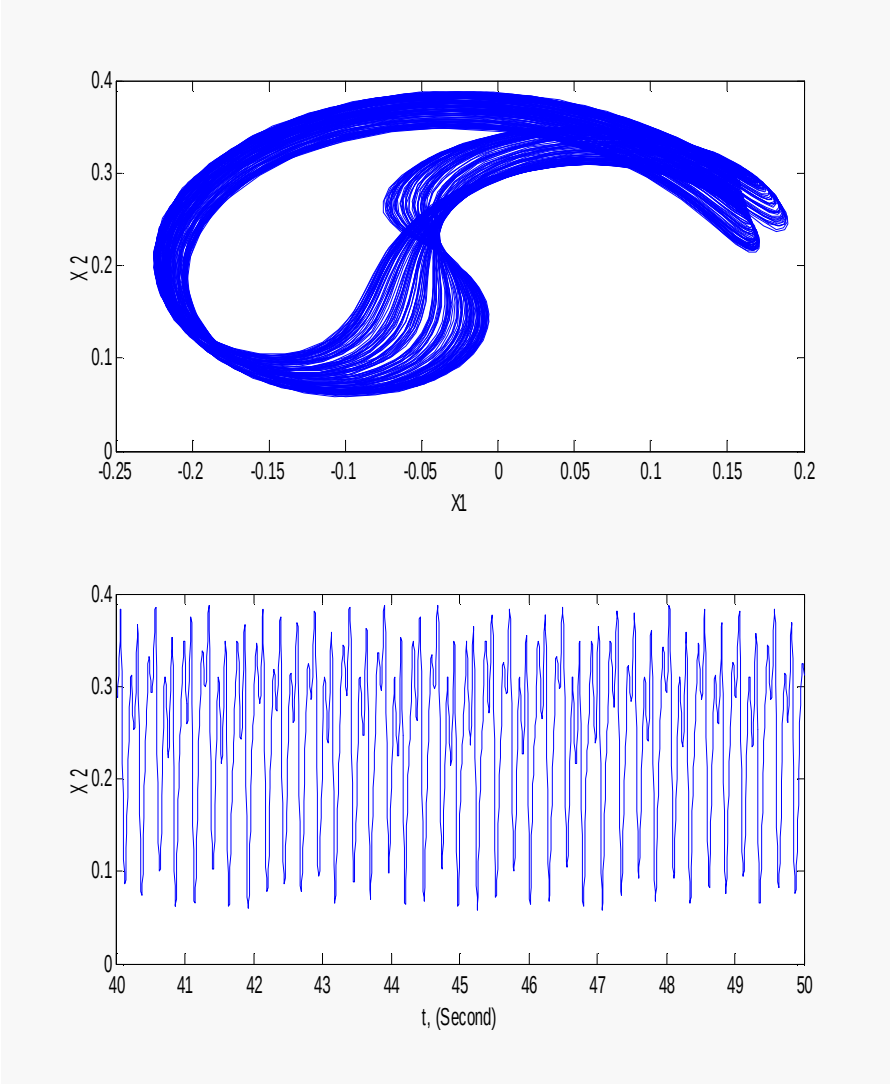

Fig:5 Phase and time plot showing Chaotic orbits at K=3.15 and $T_l$=0.5

## V. Bifurcation Behavior of the System

A bifurcation diagram shows the long-term qualitative change (equilibria/fixed points or periodic orbits) of a system as a function of a bifurcation parameter of the system. The complete dynamics of the system with the variation of the parameter can be shown with the help of bifurcation diagram. To know the complete dynamics of the system, K (the ratio of rotor time constant to its estimate) is taken as bifurcation parameter.

The Hopf bifurcation:

The Fig. 6 (a) and (b) show the phase and time plot of the IFOC induction motor at K=1.1 and load torque 0.2 Nm. It is evident from the figure that the trajectories are converging to a fixed point. The co-ordinates of the fixed point are (-0.0023, 0.4534, 0.0 0.2040). The eigenvalues of the Jacobian matrix at the fixed point are 0.5264+i27.8839, -0.5264-i27.8839, -14.2910+i0.4546, -14.2910-i0.4546.. A hopf bifurcation occurs when the value of K is increased to 1.2 A hopf bifurcation corresponds to the situation when parameter K passes through a critical value 1.2 and one pair of complex conjugate eigenvalues of the system Jacobian matrix moves from the left-half plane to right half , crossing the imaginary axis, while all other eigenvalues remain stable. At the moment of crossing, the real parts of the two eigenvalues become zero and the stability of the existing equilibrium changes from being stable to unstable. Also at the moment of crossing, a limit cycle is born. At this critical value, the eigen values are i28.2072, -i28.2072, -15.5299, -14.0929.

The eigen values of the Jacobian matrix at the corresponding equilibrium points can be computed for each value of K, giving the locus of the eigen values shown in Fig. 7. In this figure only two branches are plotted since other two branches are far away to the left and do not contribute to this bifurcation. The critical value of K can be found from the locus.

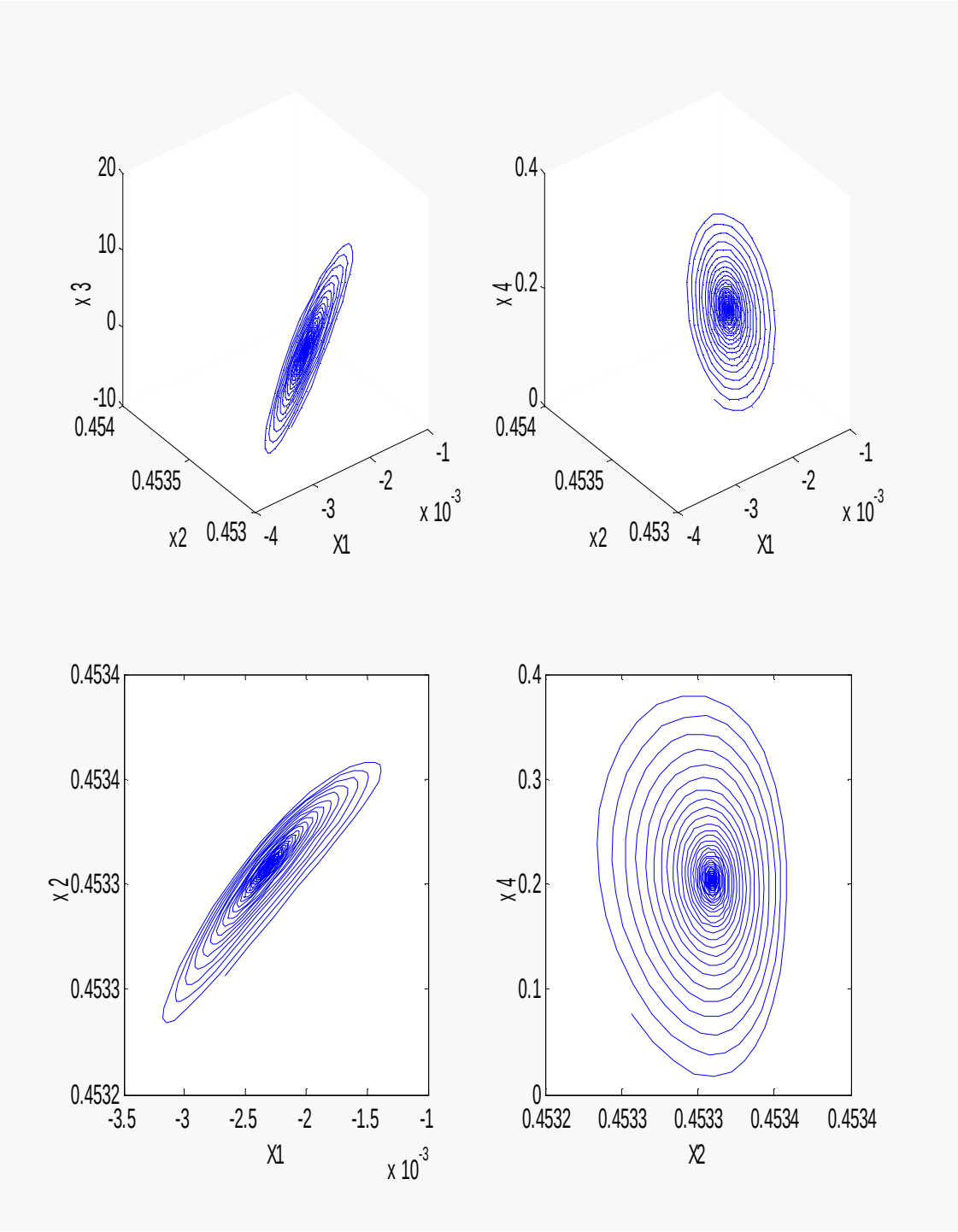

(a)

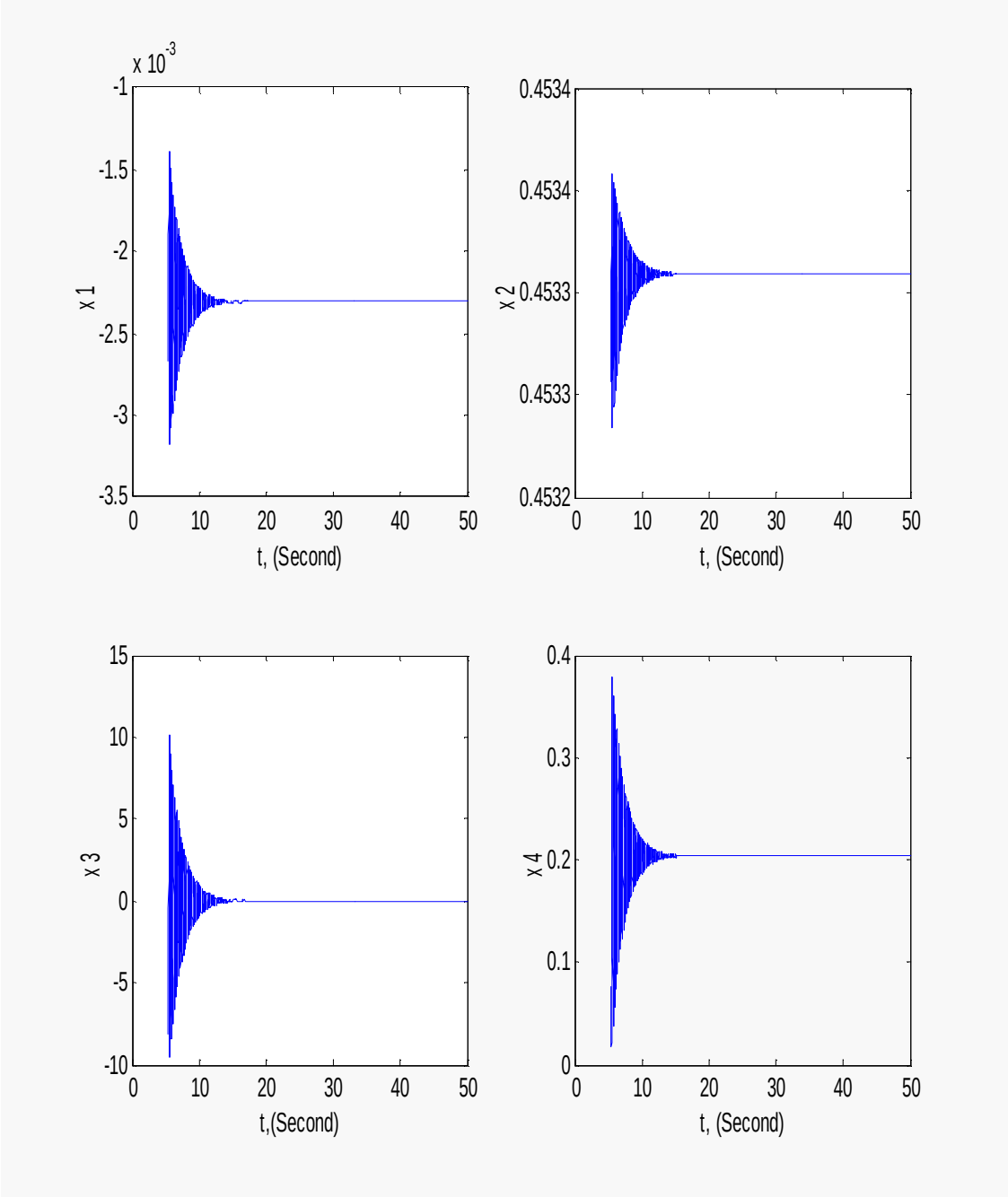

.

(b)

Fig. 6. (a) Phase plot with K=1.1 and Tl=0.2

(b) Time plot with K=1.1 and Tl=0.2

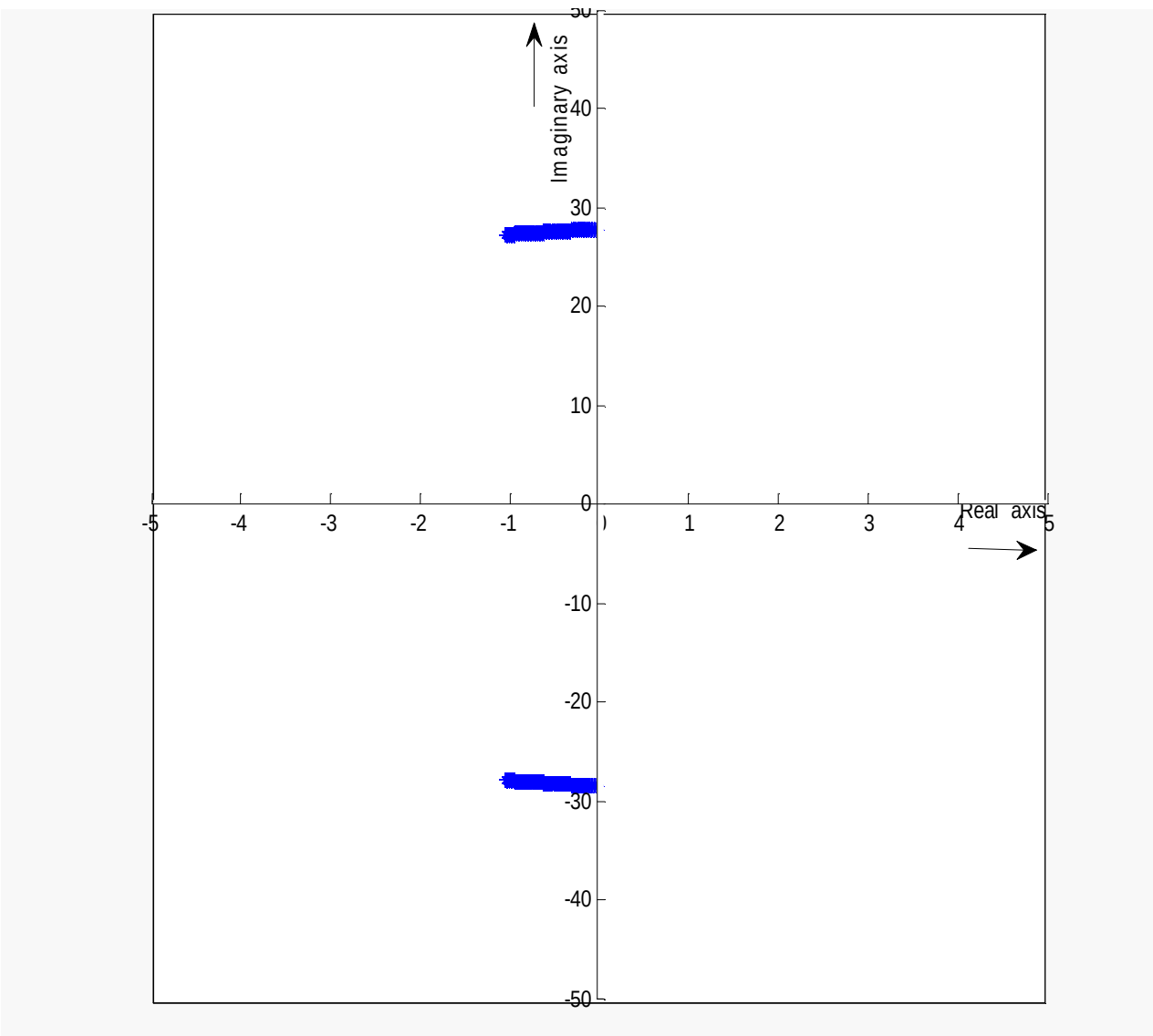

Fig. 7. Locus of the eigen values of Jacobian matrix

The oscillatory behavior of the IFOC motor is observed when the value of K is increased beyond the critical value. The phase plot of different state variables and time plot of $x_2$ and $x_4$ are shown in Fig. 8. It is evident from the Fig. 7 that the motor's behavior changes to oscillatory from fixed point after a phase of transients that last approximately up to 92 seconds.

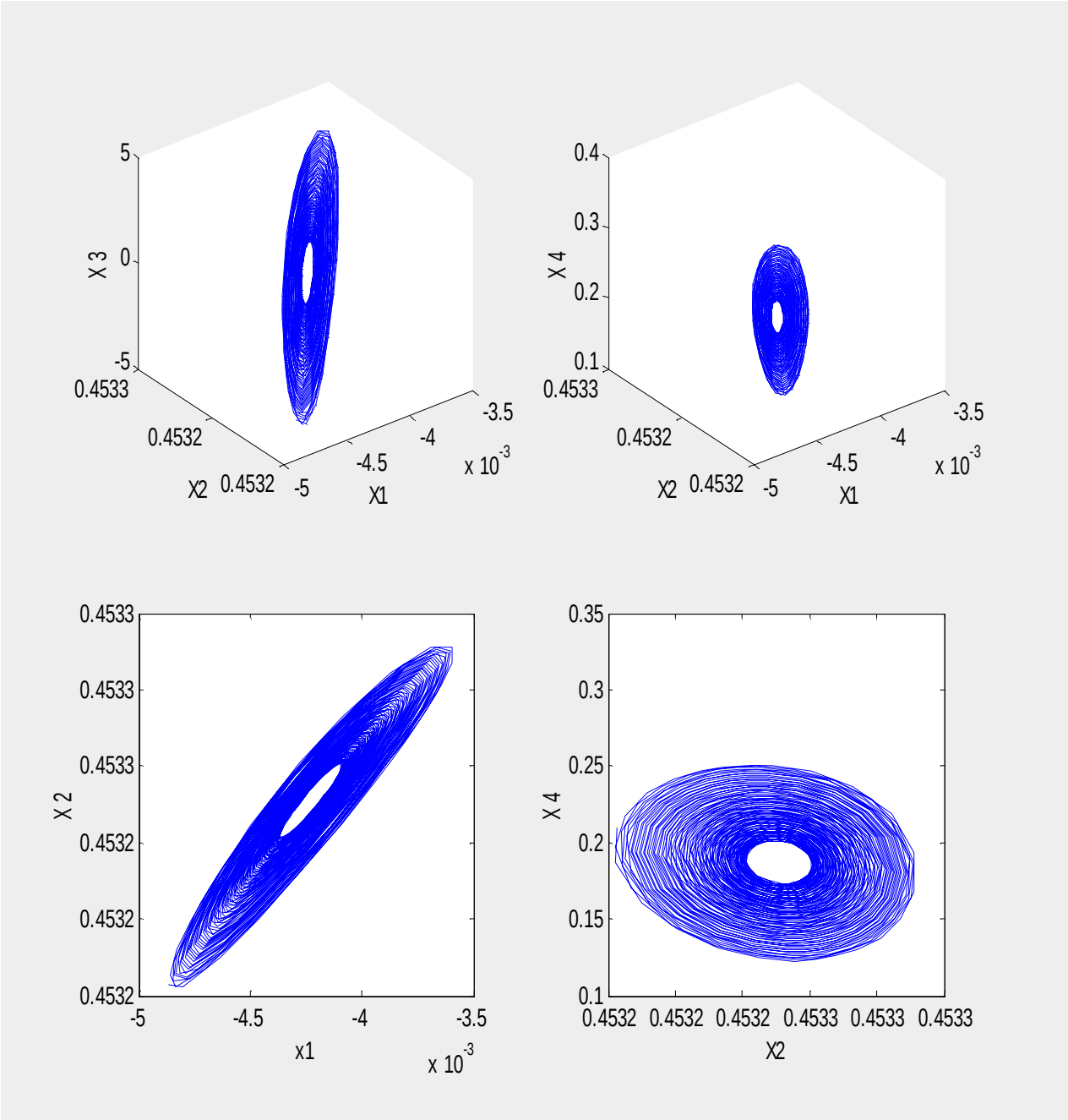

(a)

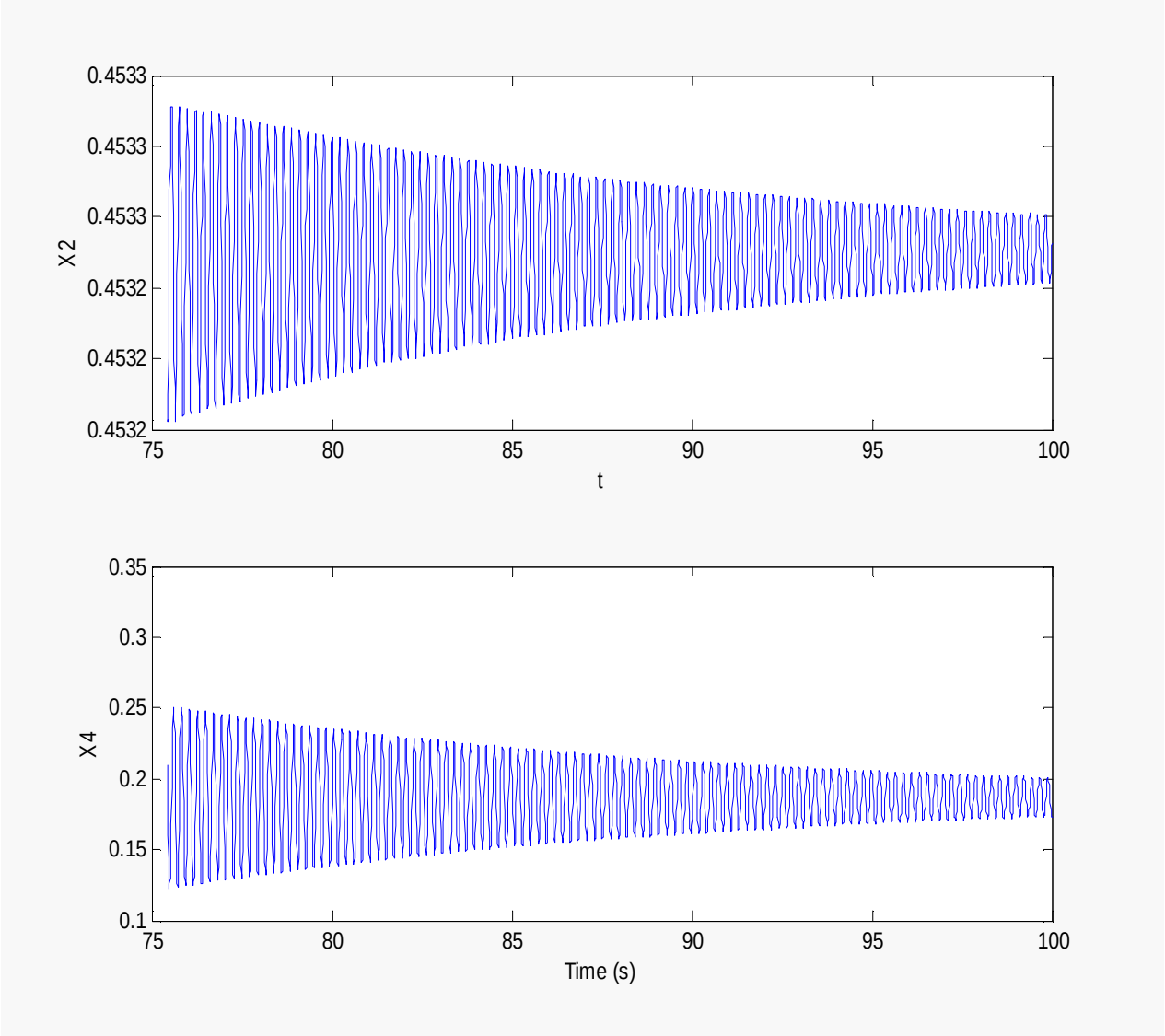

(b)

Fig.8. (a) Phase plot with K=1.205 and Tl=0.2

(b) Time plot with K=1.205 and Tl=0.2

In order to illustrate the application indices to identify Hopf bifurcation, the constant K is increased from 1.0 at a step size of 0.001 and load torque of 0.2 Nm. The index1 and index2 are calculated at each value of the constant K . The plot of index1 and index2 are shown in Fig. 9. and Fig.10. All the results including minimum singular value and Schur's decomposition are computed in MATLAB 7.5.0.

It is observed from the Fig.9 that with the increase of k, the value of the indicex1 decreases linearly and becomes zero at the bifurcation point.

Fig. 10 shows the variation of the index2 with constant k. Here the variation is not linear. The value of index2 reaches to a minimum value at the bifurcation point. Ideally this value should be zero at the bifurcation point. Some aspects like tolerances of Schur's decomposition influences indices not reaching zero at bifurcation point. The critical value of K found from the locus of eigen values of Jacobian matrix of IFOC motor and critical value of K found from the plot of index1 and index 2 are in good agreement.

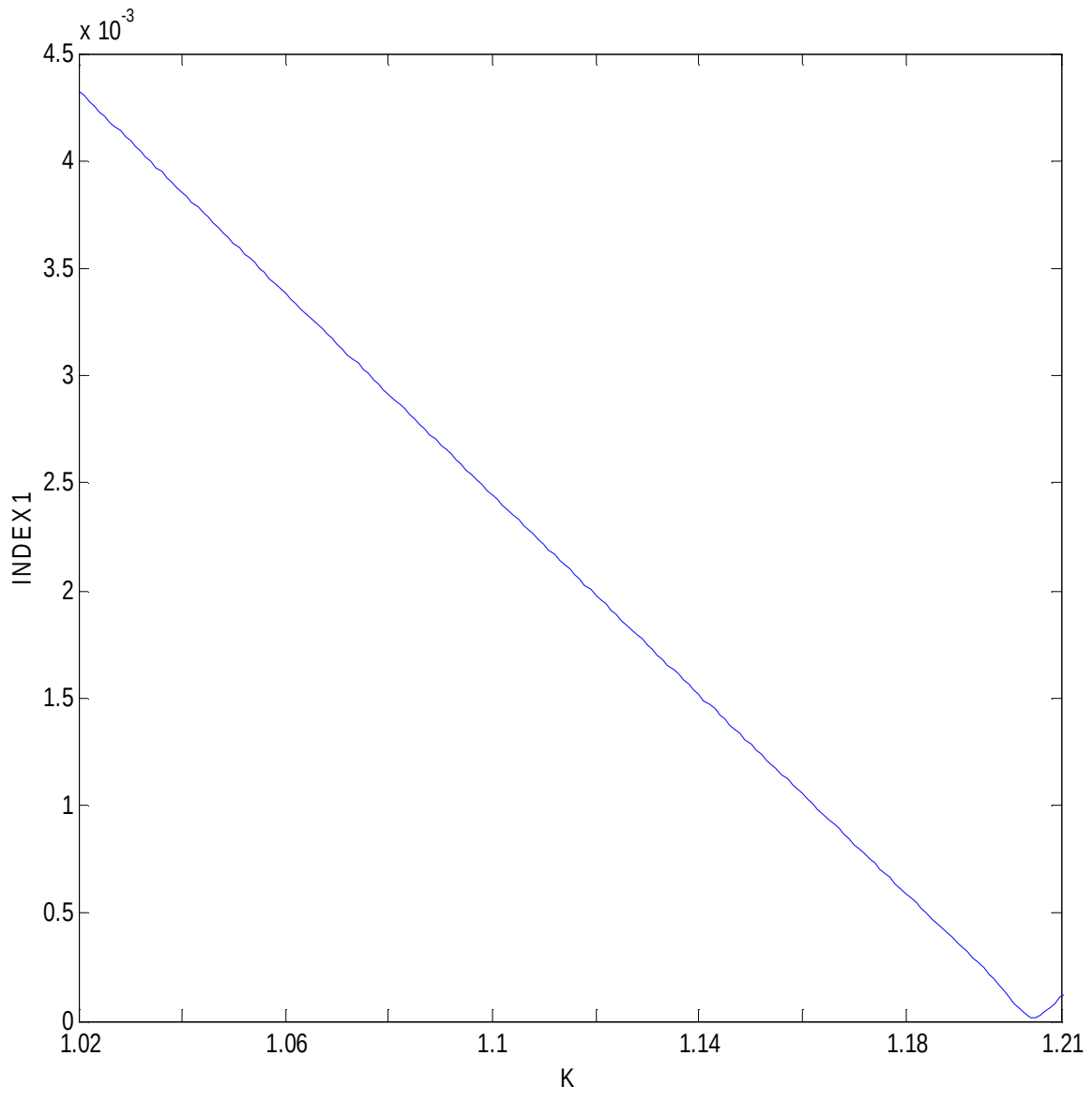

Fig. 9. Plot of index 1 with variation of K along x axis.

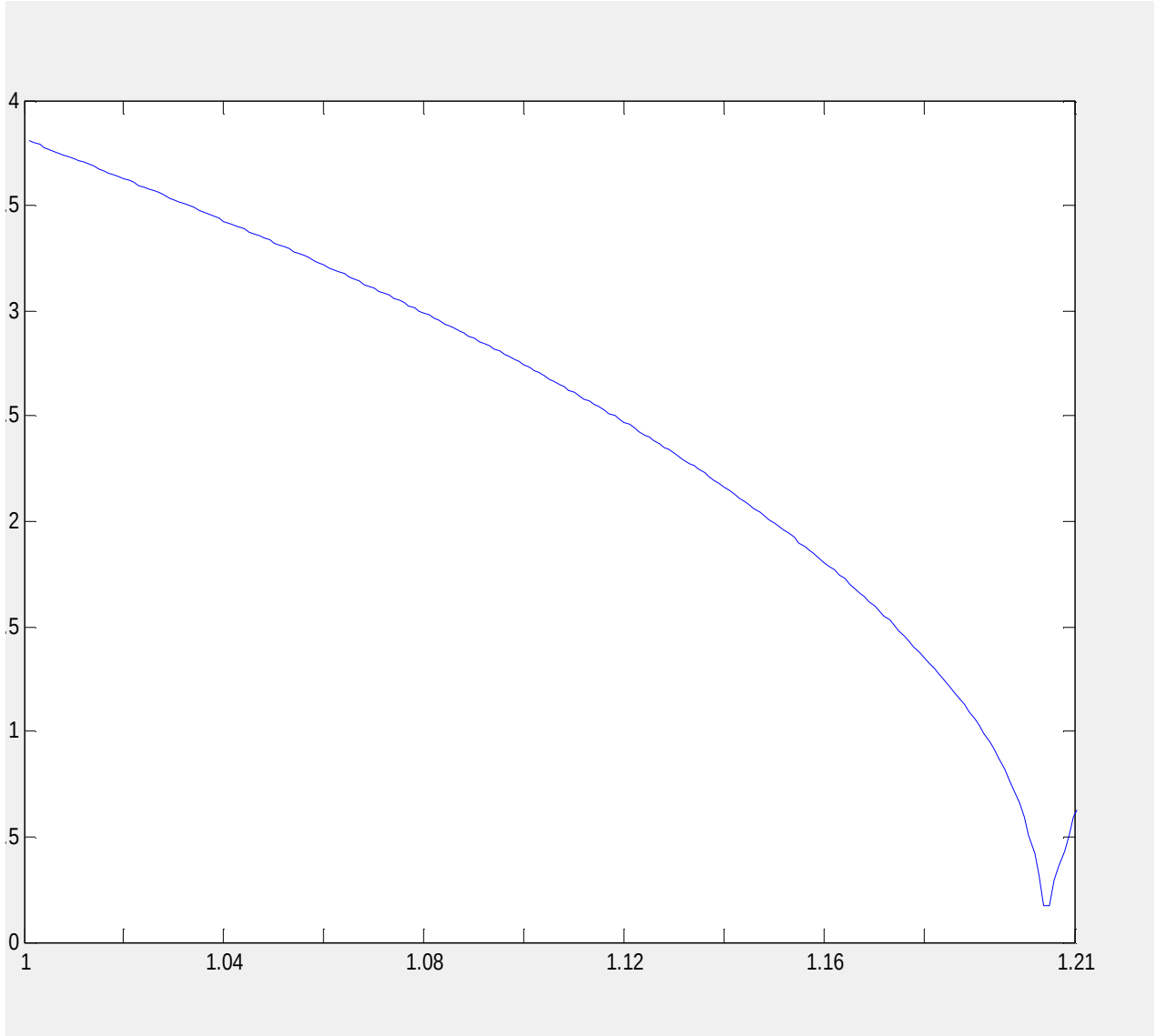

Fig.10 Plot of index 2 with variation of K along x axis.

## VI. CONCLUSION

In order to illustrate the effect of variation of parameters on the system's performance, we have presented bifurcation analysis of IFOC induction motor. It is also shown that the IFOC motors are prone to instability due to Hopf bifurcation. It is very essential to know the critical value of the parameter at which the instability arising out of oscillation occurs due to Hopf bifurcation. To identify, the critical value, two indices are presented. It is shown that these indices are capable of identifying Hopf bifurcation. The index1 shows a linear behavior with respect to changes of constant K, helping the operator of the machine to predict the proximity to the oscillation of the motor behavior due to Hopf bifurcation.

But it needs more computational efforts due to rise of problem dimension from N to 2N. Through the variation of index2 are not linear with respect to change of the constant K, it requires less computational effort  to compute the index.  However the limitation of index2 not reaching zero at the bifurcation points need to be studied further.  Those indices can also be used to determine the occurrence of other type of bifurcation.